\documentstyle[11pt,AATS,epsf]{article}	
\def\gtaprx{ \mathrel{  \vcenter{
                        \offinterlineskip \hbox{$>$}
                        \kern 0.3ex \hbox{$\sim$}    } } }
\def\ltaprx{ \mathrel{  \vcenter{
                        \offinterlineskip \hbox{$<$}
                        \kern 0.3ex \hbox{$\sim$}    } } }
\begin{document}	

\title{Nucleosynthesis Clocks and the Age of the Galaxy} 


\author{James W. Truran}
\affil{Department of Astronomy \& Astrophysics, Enrico Fermi Institute, 
University of Chicago, Chicago, IL 60637}

\author{Scott Burles}
\affil{Department of Astronomy \& Astrophysics, University of Chicago, 
Chicago, IL 60637}

\author{John J. Cowan}
\affil{Department of Physics \& Astronomy, University of Oklahoma,
Norman OK 73019}

\author{Christopher Sneden}
\affil{Department of Astronomy, University of Texas, Austin, TX 78712}


\begin{abstract}

Nucleocosmochronology involves the use of the abundances of radioactive 
nuclear species and their radiogenic decay daughters to establish the finite 
age of the elements and the time scale for their formation. While there 
exist radioactive products of several specific nucleosynthesis mechanisms 
that can reveal the histories of these mechanisms, it is the long lived 
actinide isotopes $^{232}$Th, $^{235}$U, and $^{238}$U, formed in the 
{\em r}-process, that currently play the major role in setting the time scale 
of Galactic nucleosynthesis. Age determinations in the context of Galactic 
chemical evolution studies are constrained by intrinsic model uncertainties. 
Recent studies have rather taken the alternative approach of dating individual 
stars. Thorium/europium dating of field halo stars and globular cluster stars 
yields ages on the order of 15$\pm$4 Gyr. A solid lower limit on stellar ages 
is available as well, for stars for which one both knows the thorium abundance 
and has an upper limit on the uranium abundance. For the cases of the two 
extremely metal deficient field halo stars CS 22892-052 and HD 115444, lower 
limits on the nuclear age of Galactic matter lie in the range 10-11 Gyr. 
Th/U dating of the star CS 31082-00 gives an age of 12.5$\pm$3 Gyr. Observations
of thorium and uranium abundances in globular cluster stars should make possible
nuclear determinations of the ages of clusters that can be compared with ages 
from conventional stellar evolution considerations. 

\end{abstract}

\section{Introduction}


Nuclear chronometers have played an historically important role in the 
determination of the ages of our solar system and our Milky Way Galaxy, 
thus providing important lower limits on the age of the Universe itself. 
The critical long lived nuclear radioactivities were long ago identified 
to be $^{187}$Re ($\tau_{1/2}$=4.5x10$^{10}$ years), $^{232}$Th 
($\tau_{1/2}$=1.4x10$^9$), $^{235}$U ($\tau_{1/2}$=7.0x10$^8$), and 
$^{238}$U ($\tau_{1/2}$=4.5x10$^9$), all of which are products of the 
{\em r}-process of neutron capture synthesis {Burbidge {\it et al.} 1957; 
Cameron 1957). 

While interesting age constraints can be obtained on the basis of "model 
independent" chronometric considerations (see, e.g. Meyer \& Schramm 1986), 
most determinations of the age of the Galaxy to date have been obtained in the 
context of models of Galactic chemical evolution (see, e.g. the review by 
Cowan, Thielemann, \& Truran 1991a). Such models can be used to explore the 
sensitivities of the ages thus obtained to the star formation and 
nucleosynthesis history of Galactic matter. The Re/Os chronometer is of 
particular concern in this regard, both because the $^{187}$Re decay rate is 
temperature sensitive (a factor when $^{187}$Re is recycled through stars) 
and because both r-process and s-process contributions may be important. 

In this paper we will concentrate on recent age determinations for individual 
stars, built upon the use of thorium and uranium related chronometers. Since 
the available data is associated with extremely metal-poor stars, formed 
during the very earliest stages of Galactic evolution, dating of such stars 
is entirely equivalent to dating of the Galaxy as a whole.

\section{r-Process Nucleosynthesis Considerations} 

While the general nature of the r-process of neutron capture synthesis and 
its contributions to the abundances of heavy elements in the mass range from 
beyond the iron peak through thorium and uranium is generally understood 
(Cowan, Thielemann, \& Truran 1991b), many of the details remain to be worked 
out. Of particular concern is the very fact that the astrophysical site in which
the r-process occurs remains unidentified. Both meteoritic data (Wasserburg, 
Busso, \& Gallino 1996) and stellar abundance data (e.g. the robust consistency 
of the heavy (A $\gtaprx$ 140) r-process abundance patterns in the most 
extreme metal deficient stars with that of solar system matter) suggest the 
likelihood of two distinct r-process environments responsible, respectively, 
for the mass regions A $\ltaprx$ 140 and A $\gtaprx$ 140. The heavy element 
patterns in low metallicity stars strongly imply that at least the production of
r-process nuclei in the heavy element region is associated with stars of short 
lifetimes (e.g. massive stars and/or supernova Type II). 

Excellent reviews of r-process nucleosynthesis through the years have been 
provided by Hillebrandt (1978), Cowan, Thielemann \& Truran (1991b), and 
Meyer (1994). Promising studied sites for the synthesis of the r-process 
isotopes in the mass range A $\gtaprx$ 140 include: (1) the high-entropy, 
neutrino heated, hot bubble associated with Type II supernovae (Woosley 
{\it et al.} 1994; Takahashi, Witti, \& Janka 1994); (2) the ejection and 
decompression of matter from neutron star-neutron star mergers 
(Lattimer {\it et al.} 1977; Freiburghaus {\it et al.} 1999); and (3) the ejection
of neutronized matter in magnetic jets from collapsing stellar cores 
(LeBlanc \& Wilson 1970). Note that all three of these mechanisms are tied to 
environments provided by the evolution of massive stars (and associated 
Type II supernovae) of short lifetimes ($\tau$ $\ltaprx$ 10$^8$ years), 
compatible with their presence in the oldest stellar populations of our 
Galaxy. 

%
%
%
%

\section{Dating Individual Stars}

The direct determination of the ages of individual stars has recently become 
possible, with the availability of thorium (and most recently uranium) 
abundances for halo stars. A distinct advantage of this approach is the 
elimination of uncertainties associated with our imperfect knowledge of the 
star formation, stellar evolution, and associated nucleosynthesis histories 
of galaxies, which are a necessary ingredient of chemical-evolution-based 
models of nucleocosmochronology. There remain. of course, the uncertainties 
associated with the nuclear properties of heavy actinide nuclei, which dictate 
the relative levels of production of $^{232}$Th, $^{235}$U, and $^{238}$U 
in nucleosynthesis sites, and with the character of the astrophysical 
r-process itself. For example, a critical question associated with the use of 
the Th/Eu ratio as a chronometer is the robustness of the reproduction of 
the solar system r-process pattern over the mass range 
130 $\ltaprx$ A $\ltaprx$ 238. We should also note that the use of such age 
determinations for individual stars (particularly the Th/U chronometer) may 
be {\it effectively} constrained to low metallicity ([Fe/H] $\ltaprx$ -2.5), 
r-process enriched ([r-process/Fe] $\gtaprx$ 1) stars like CS 22892-052, 
CS 31082-001, and HD 115444. 

\subsection{Th/Eu Dating}

The detection of thorium and the determination of its abundance in the 
extremely metal deficient halo field star CS 22892-052 first made thorium 
dating possible (Sneden {\it et al.} 1996). In this case, due to the absence 
of knowledge of the uranium abundance, it is necessary to utilize some 
stable r-process product. The choice of europium here is a relatively 
obvious one, since both isotopes of Eu are produced almost exclusively by 
the r-process. 

Data regarding Th and Eu is now available for a number of stars in the halo and 
in globular clusters. This data is collected in our Table 1. The ages were 
determined with the assumption that the "theoretical" r-process production 
ratio was (Th/Eu)$_{theory}$= 0.51. Note that the mean Thorium/Europium age 
for these stars is 13.6 Gyr, with a spread  
consistent with the intrinsic observational uncertainty of approximately 
$\pm$ 4 Gyr.

\begin{table}
\par\noindent
\centerline{TABLE 1}
\vskip .1truein
\centerline {Thorium/Europium Ages for Individual Stars }
$$\vbox {\tt \halign {\hfil #\hfil && \quad \hfil #\hfil \cr
\noalign{\hrule} \noalign{\vskip1pt}\cr
\noalign{\hrule} \noalign{\bigskip}\cr
 Object &  [Fe/H] & log$\epsilon_{Th}$ & log$\epsilon_{Eu}$ & $\Delta$ &
   (Th/Eu) & $\tau_*$\cr \noalign{\hrule}
\noalign{\bigskip}

\ CS22892-052  & -3.1 & -1.57 & -0.91 & -0.66 & 0.22 & 16.8   \cr
\ HD 115444    & -3.0 & -2.23 & -1.63 & -0.60 & 0.25 & 14.4   \cr

\ & & & & &  \cr
\ HD 115444    & -3.0 & -2.21 & -1.66 & -0.55 & 0.28 & 12.1   \cr
\ HD 186478    & -2.6 & -2.25 & -1.55 & -0.70 & 0.20 & 18.9   \cr
\ HD 108577    & -2.4 & -1.99 & -1.48 & -0.51 & 0.31 & 10.1   \cr
\ M92 VII-18   & -2.3 & -1.94 & -1.45 & -0.49 & 0.32 &  9.4   \cr
\ BD +8 2856   & -2.1 & -1.66 & -1.66 & -1.16 & 0.32 &  9.4   \cr
\ & & & & &  \cr

\ K341 (M15)   & -2.4 & -1.47 & -0.88 & -0.59 & 0.25 & 14.4   \cr
\ K462 (M15)   & -2.4 & -1.26 & -0.61 & -0.65 & 0.22 & 16.8   \cr
\ & & & & &  \cr

\ CS31082-001   & -3.0 & -0.96 & -0.70 & -0.26 & 0.55 &  -     \cr

\ & & & & &  \cr

\    & \cr \noalign{\hrule} \noalign{\bigskip}}}$$

\medskip\noindent
References: (1) Sneden {\it et al.} 2000a; (2) Westin {\it et al.} 2000;
(3) Johnson \& Bolte 2001; (4) Sneden {\it et al.} 2000b; Hill {\it et al.}
2001.

\end{table}

A critical question concerning the use of the Th/Eu chronometer is the 
robustness of the r-process abundance pattern over a range 
of mass number A $\approx$ 140-238. The abundance patterns in this mass number 
range for the two extremely metal deficient stars CS 22892-052 and HD 115444, 
displayed in Figure 1, both show a remarkable agreement with the solar 
system r-process abundance distribution. The quoted abundance level for 
the star CS 31082-001 (Hill {\it et al.} 2001), however, is distinctly 
different from those for the other stars in  Table 1 (where we do not quote a
Th/Eu age for this star). It is important to 
examine this behavior for other halo and globular cluster stars to 
more firmly establish the limits of confidence of the Th/Eu chronometer. 
We emphasize that we believe that the Th/Eu chronometer (or perhaps a 
Th/Pt chronometer) will generally be a more commonly available tool for 
the dating of halo stars than Th/U - and thus that we should work hard to 
confirm its reliability. 

\begin{figure}[htb]
\plotone{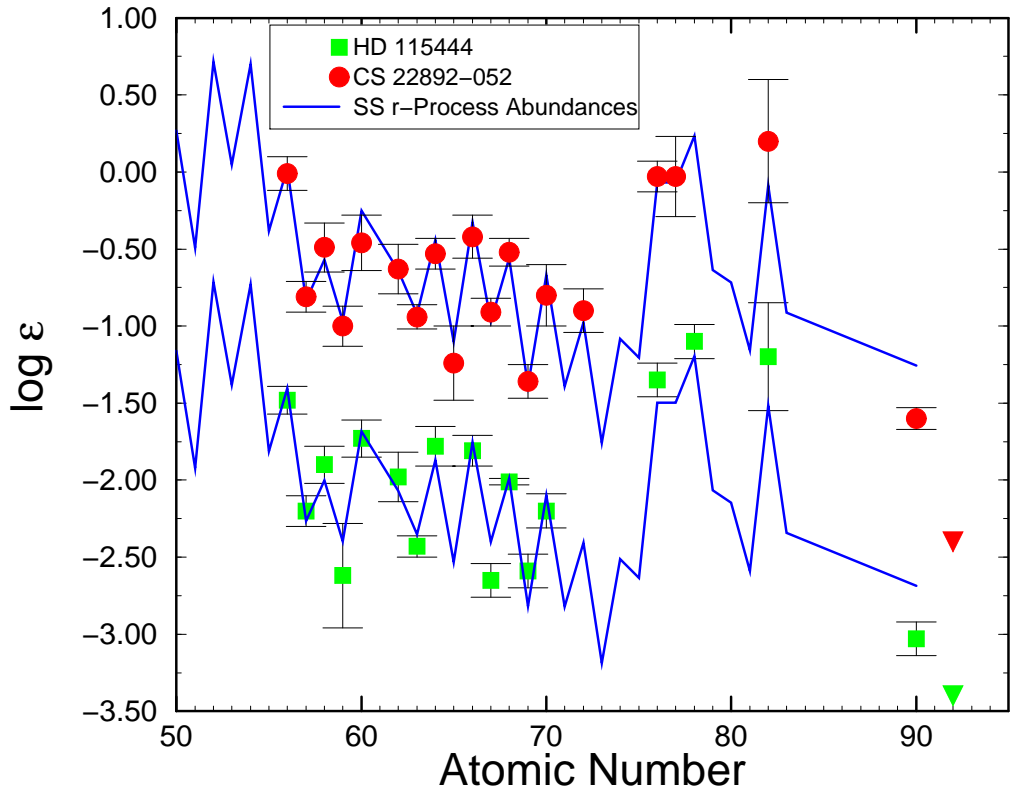}
\caption{
Abundance comparisons between HD 115444 (filled squares) and
CS 22892--052 (filled circles). The abundance values for HD 115444 have been
vertically displaced (arbitrarily by -0.8) for display purposes.
The upside-down triangles are upper limits on the uranium abundances.
The solid lines are scaled solar system
r-process only abundance  distributions.
\label{fig1}}
\end{figure}

\subsection{Th/U Dating}

When available, it is understood that the Th/U chronometer will give age 
determinations for which the nuclear uncertainties and astrophysical 
uncertainties may be smaller than those for the Th/Eu chronometer. 
Theoretical calculations of r-process nucleosynthesis can be employed to 
provide estimates of for example the $^{232}$Th/$^{238}$U and 
$^{235}$U/$^{238}$U ratios. Early model predictions for these so-called 
production ratios typically show rather wide dispersion (Cowan, Thielemann, 
\& Truran 1987). However, more recent theoretical determinations 
(e.g. M\"oller, Nix, \& Kratz 1997; see also Cowan {\it et al.} 1999) 
are now available. 

Recently Cayrel et~al. (2001; see also their contribution to this
conference volume) have discovered that the ultra metal poor star CS~31082-001
([Fe/H]~$\sim$~--3) has even larger $n$-capture abundance levels
than does CS~22892-052.
Most interesting is their detection of $many$ transitions of \ion{Th}{2} and
the strongest \ion{U}{2} line; the combined abundances of two
radioactive elements (with different half-lives) yields an  
``age'' of the neutron-capture elements in this star of $\sim$13~Gyr.
But they also report an $observed$ [Th/Eu] ratio that is much larger
than in CS~22892-052. 
This abundance ratio in CS~31082-001 would imply an age of only a few Gyrs,
obviously inconsistent with this star's metallicity and halo membership 
(see, e.g. the data cited in our Table 1). 
It is clear that many more Eu and Th abundances need to be determined
for halo stars, in order to make some sense of this issue.

We also wish to note, however, the 
{\it even upper limits on the uranium abundance} allow firm and interesting 
lower limits on the age of the Galaxy. Figure 2 displays the exponential 
dependence of the age of a star as a function of the observed Th/U ratio.
Here we assume that $^{238}$U dominates $^{235}$U, and therefore show only ages 
greater than 5 Gyrs. The green region indicates the range of allowed 
r-process Th/Eu production ratios (Cowan, Thielemann, \& Truran 1987). 
Lines of constant log(U/Th) are displayed. An upper limit on U/Th excludes 
the lower left half of the plane. We note that the upper limits on the 
uranium abundances for the two stars CS~22892-052 and HD 115444 allow lower 
limits on the ages of these stars, respectively, of 10 and 11 Gyr. 

\begin{figure}[htb]
\plotone{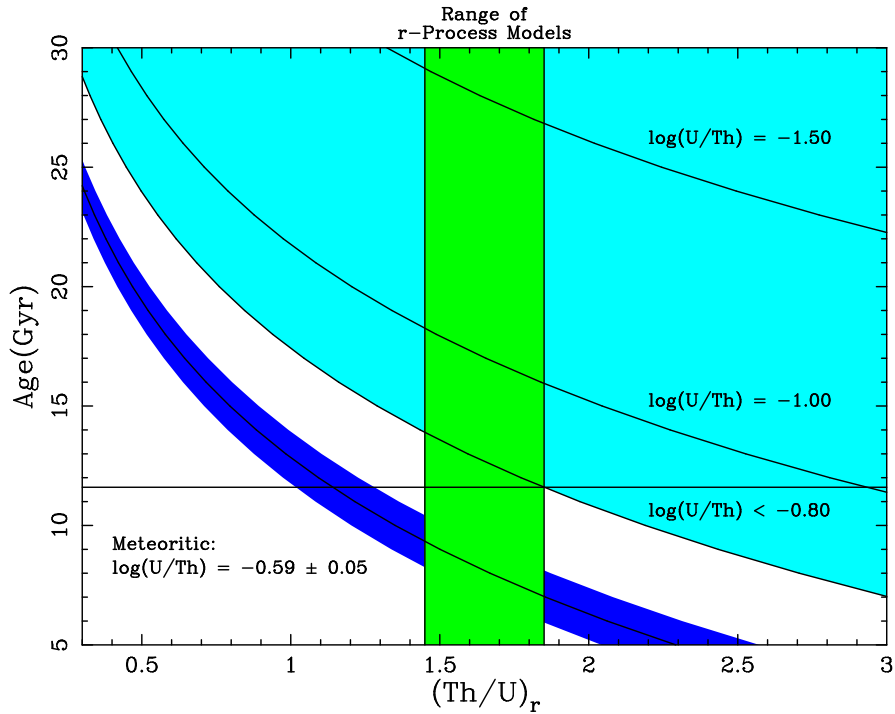}
\vskip .2truein
\caption{
Stellar ages as a function of the Th/U ratio.
Shown is the exponential dependence of the observed Th/U abundance
on the age of the system.  We assume that 238 dominates 235, and
therefore only show ages greater than 5 Gyrs.  Lines of constant
log U/Th ratios are shown in the plane of Age vs. 232/238 seed ratio.
An upper limit on U/Th excludes the lower left half of the plane.
A measurement of U/Th is restricted to a narrow band as is the case
with meteoritic observations (Anders \& Grevesse 1989).
\label{fig2}}
\end{figure}

%
%

\section{Conclusions }

We draw the following conclusions from our overview of considerations of 
nuclear chronometer dating: 

\begin{itemize}

\item{(1)} The important nuclear chronometers $^{232}$Th, $^{235}$U, 
and $^{238}$U  
are r-process products.

\item{(2)} The observations of r-process abundance patterns in the oldest 
([Fe/H] $\ltaprx$ -2.5) halo stars serve  
to confirm the identification of the r-process site with massive star environments 
and to convince us that we are dating the history of stellar activity in our Galaxy. 

\item{(3)} The robustness of the r-process mechanism for 
the production of nuclei 
in the region A $\gtaprx$ 140 is reflected in the extraordinary agreement 
(see Figure 1) of the 
stellar abundance patterns with the solar system r-process abundances (even at 
metallicities so low that at best only a few supernovae can have contributed). 
{\it The remarkable star CS 22892-052, at [Fe/H] = -3.1, exhibits a pure r-process 
pattern in the mass range  A $\gtaprx$ 140, but at an abundance level 
[r-process/Fe] $\approx$ +1.5.}



\item{(4)} Ages (and/or age constraints) for individual halo field stars or 
globular cluster stars can be obtained (in principle) from Th/Eu and Th/U, 
when the abundances of these nuclear species are known. 

\item{$\rightarrow$} Th/Eu dating of the two 
well studied stars 
CS 22892-052 and HD 115444 give an average age chronometric age 
$\tau_*$ $\sim$ 15.6 $\pm$ 4.6 Gyr. 

\item{$\rightarrow$}  Upper limits on the Uranium abundances for 
these two stars provide lower limits on 
their ages $\tau_*$ $\gtaprx$ 10-11 Gyr.

\item{(5)} While both thorium and uranium abundance determinations for 
individual halo stars will continue to accumulate, it seems likely that 
uranium abundances will become available for only a small fraction of those for 
which thorium abundances are known. Continued efforts to place the Th/Eu 
(or perhaps Th/Pt) chronometer on a firm basis are thus critical to our 
ability to date the various components of our Galactic halo population. 

\end{itemize}

\acknowledgments  
Our work on $n$-capture elements in halo stars has been a collaborative
effort over many papers, and we thank  our co-authors for their
contributions over the years. This research has been supported in part by 
NSF grant AST 9986974 to the University of Oklahoma (JJC) and by the 
ASCI/Alliances Center for Astrophysical Thermonuclear Flashes at the 
University of Chicago under DOE grant B341495 (JWT). 


\par\vfil\eject


\begin{references}

\reference Anders, E., \& Grevesse, N. 1989, Geochim. Cosmochim. Acta., 46, 2363

\reference Burbidge, E.M., Burbidge, G.R., Fowler, W.M., \& Hoyle, F. 
1957, RMP, 29, 547.

\reference Burris, D. L., Pilachowski, C. A., Armandroff, T. A.,
Sneden, C., Cowan, J. J., \& Roe, H. 2000, \apj, 544, 302. 

\reference Cameron, A.G.W. 1957, Chalk River Report CRL-41.

\reference Cayrel, R., et al.  
2001, Nature, 409, 691. 

\reference Cowan, J. J., Pfeiffer, B., Kratz, K.-L., Thielemann, F.-K.,
Sneden, C., Burles, S., Tytler, D., \& Beers, T. C. 1999, \apj, 521, 194. 

\reference Cowan, J.J., Thielemann, F.-K., \& Truran, J.W. 1987, 
ApJ, 323, 543. 

\reference Cowan, J.J., Thielemann, F.-K., \& Truran, J.W. 1991a, 
ARAA, 29, 447.

\reference Cowan, J.J., Thielemann, F.-K., \& Truran, J.W. 1991b, 
Phys. Reports, i208, 267. 

\reference Freiburghaus, C., Rembges, J.-F., Rauscher, T., Kolbe, E., 
Thielemann, F.-K., Kratz, K.-L., Pfeiffer, B., \& Cowan, J.J. 1999, 
ApJ, 516, 381.

\reference Hill, V., Plez, B., Cayrel, R., \& Beers, T.C. 2001, 
these proceedings.

\reference Hillebrandt, W> 1978, Sp. Sci. Rev., 21, 639.

\reference Johnson, J.A.. \& Bolte, M.S. 2001, ApJ, in press.

\reference Lattimer, J.M., Mackie, F., Ravenhall, D.N., \& Schramm, D.N. 1977, 
ApJ, 213, 225. 

\reference LeBlanc, J.M., \& Wilson, J.R. 1970, ApJ, 161, 541.

\reference Meyer, B.S. 1994, ARAA, 32, 153. 

\reference Meyer, B.S., \& Schramm, D.N. 1986, ApJ, 311, 406.

\reference M\"oller, P., Nix, J.R., \& Kratz, K.-L. 1997, 
At. Data Nucl. Data Tables, 66, 131. 

\reference Sneden, C., Cowan, J.J., Ivans, I.I., Fuller, G.M., Burles, S., 
Beers, T.C., \& Lawler, J.E. 2000a, ApJ, 533, L139.

\reference Sneden, C., Johnson, J., Kraft, R. P., Smith, G. H., Cowan, J. J., 
Bolte, M. S. 2000b, \apj, 536, L85. 

\reference Sneden, C., McWilliam, A., Preston, G. W., Cowan, J. J.,
Burris, D. L., \& Armosky, B. J. 1996, \apj, 467, 819.

\reference Takahashi, K., Witti, J., \& Janka, H.-T. 1994, A\&A, 286, 857.

\reference Toenjes, R. {\it et al.} 2001, these proceedings.

\reference Truran, J. W. 1981, \aap, 97, 391.

\reference Wasserburg, G. J., Busso, M., \& Gallino, R. 1996, \apj, 466, L109.

\reference Westin, J., Sneden, C., Gustafsson, B., \& Cowan, J. J. 2000,
\apj, 530, 783.

\reference Woosley, S.E., Wilson, J.R., Mathews, G.J., Hoffman, R.D., \& 
Meyer, B.S. 1994, ApJ, 433, 229.

\end{references}
\end{document}